\documentclass[aps,pre,reprint,twocolumn,groupedaddress,nofootinbib]{revtex4-2}

\usepackage{graphicx}
\usepackage{bm}
\usepackage{amsmath}
\usepackage{amssymb}
\usepackage{amsfonts}
\usepackage{amsthm}
\usepackage{hyperref}
\usepackage[dvipsnames]{xcolor}

\definecolor{crimson}{RGB}{165,28,48}
\hypersetup{colorlinks=true,allcolors=crimson}

\DeclareMathOperator{\tr}{tr}

\renewcommand{\Re}{\operatorname{Re}}
\renewcommand{\Im}{\operatorname{Im}}

\begin{document}
\title{Stationary covariance spectra of discrete-time non-normal random recurrent dynamics}

\author{Jacob A. Zavatone-Veth}
\email{jzavatoneveth@rockefeller.edu}
\altaffiliation{Present address: Laboratory of Theoretical Neuroscience, The Rockefeller University, New York, NY, USA 10065}
\affiliation{Center for Brain Science, Harvard University, Cambridge, MA, USA 02138}
\affiliation{Society of Fellows, Harvard University, Cambridge, MA, USA 02138}

\date{\today}

\begin{abstract}
    Principal component analysis is widely used to characterize structure in the dynamics of recurrent neural networks. For stationary noise-driven dynamics, the distribution of variance among the principal components is determined by the spectrum of the stationary covariance matrix. While the spectral properties of this matrix are well-understood for linear networks with normal synaptic weight matrices, our understanding of the stationary covariance spectrum for random non-normal dynamics remains incomplete. In this note, we use a free-probability approach to show almost-sure convergence of the stationary covariance spectrum of discrete-time dynamics with random non-normal Gaussian weights, and derive a closed functional equation characterizing its distribution. This characterization allows us to analyze the behavior of tail eigenvalues in the critical regime: they decay with the inverse-square of their rank. In contrast, applying the same approach to the analogous continuous-time dynamics leads to an infinite hierarchy of Schwinger-Dyson equations, rather than a closed scalar equation. 
\end{abstract}

\maketitle

\section{Introduction}

Consider a noise-driven linear recurrent neural network, with $N$ neurons that are connected through a weight matrix $J$. There are two classes of such dynamics which one might naturally consider: the discrete-time evolution 
\begin{equation} \label{eqn:discrete}
    \mathbf{x}_{t+1} = \mathbf{J} \mathbf{x}_{t} + \mathbf{z}_t,
\end{equation}
where $\mathbf{z}_t \sim \mathcal{N}(0,I_N)$, and the continuous-time evolution
\begin{equation} \label{eqn:continuous}
    \frac{d}{dt} \mathbf{x}(t) = - \mathbf{x}(t) + \mathbf{J} \mathbf{x}(t) + \mathbf{z}(t) ,
\end{equation}
where now $\mathbf{z}(t)$ is white Gaussian noise. If $\mathbf{J}$ has i.i.d. Gaussian elements $J_{ij} \sim \mathcal{N}(0,g/N)$ for a gain parameter $g>0$\footnote{Our convention is that the \emph{variance} of the elements of $\sqrt{N}\mathbf{J}$ is $g$. We note that it is standard in the neuroscience literature to use $g$ instead for the standard deviation, and thus $g^2$ for the variance \cite{sompolinsky1988chaos,hu2022spectrum,hennequin2012nonnormal}. Because the variance appears much more frequently in our computations, directly denoting it by $g$ lightens the notation.}---that is, it is a real Ginibre matrix---then \eqref{eqn:discrete} and \eqref{eqn:continuous} constitute minimal models for non-normal recurrent dynamics in discrete or continuous time, respectively; they are the linearized equivalents of the canonical model of random recurrent dynamics \cite{sompolinsky1988chaos,toyoizumi2011beyond,hu2022spectrum,mastrogiuseppe2025stochastic,bordelon2026dynamics,hennequin2012nonnormal,chalker1998eigenvector,tian2026diversity,couillet2016echo}. 

As $\mathbf{J}$ is stable with high probability at large $N$ provided that the gain parameter $g<1$, the distribution of activity under either discrete- or continuous-time dynamics will relax to a Gaussian stationary distribution \cite{gardiner1985handbook}. In discrete time, the stationary covariance $\mathbf{\Sigma}_{d} = \lim_{t \to \infty} \mathbb{E}[\mathbf{x}_{t}\mathbf{x}_{t}^{\top}]$ solves the discrete Lyapunov equation
\begin{equation}
    \mathbf{\Sigma}_{d} = \mathbf{I}_{N} + \mathbf{J} \mathbf{\Sigma}_{d} \mathbf{J}^{\top},
\end{equation}
and can be written as an infinite series: 
\begin{equation}
    \mathbf{\Sigma}_{d} = \sum_{k=0}^{\infty} \mathbf{J}^{k} (\mathbf{J}^{\top})^{k}.
\end{equation}
In continuous time, the stationary covariance $\mathbf{\Sigma}_{c}= \lim_{t \to \infty} \mathbb{E}[\mathbf{x}(t)\mathbf{x}(t)^{\top}]$ solves the continuous Lyapunov equation
\begin{equation}
    (\mathbf{I}_{N}-\mathbf{J}) \mathbf{\Sigma}_{c} + \mathbf{\Sigma}_{c} (\mathbf{I}_{N}-\mathbf{J})^{\top} = \mathbf{I}_{N},
\end{equation}
and can be written explicitly as 
\begin{equation}
    \mathbf{\Sigma}_{c} = \int_{0}^{\infty} e^{\mathbf{J} t} e^{\mathbf{J}^{\top} t} e^{-2t}\, dt.
\end{equation}

In either case, determining the eigenvalue spectrum of the stationary covariance is of interest because it determines the distribution of variance among the principal components of the stationary activity \cite{hu2022spectrum,mastrogiuseppe2025stochastic}. This makes these simple random linear dynamical models important null models for analyses of recurrent neural computations based on principal component analysis \cite{hu2022spectrum,pachitariu2026critical,chen2024searching,stringer2019high,tian2026diversity,mastrogiuseppe2025stochastic}. Alternatively, from a control theory perspective, the stationary covariance matrices $\mathbf{\Sigma}_d$ and $\mathbf{\Sigma}_c$ coincide with the discrete and continuous controllability Gramians, respectively. Then, the spectra are of interest because the eigenvalues determine the control costs along different dimensions \cite{preciado2016gramian}. Yet a third motivation for studying these dynamics comes from non-equilibrium statistical mechanics, where Ornstein-Uhlenbeck processes of the form \eqref{eqn:continuous} with non-normal interaction matrices give a canonical, minimal example of a non-equilibrium steady state \cite{godreche2018characterising,fyodorov2025nonorthogonal,ferreira2025random}. 

In the case where the matrix $\mathbf{J}$ is normal, the spectra of $\mathbf{\Sigma}_{d}$ and $\mathbf{\Sigma}_{c}$ are relatively straightforward to characterize because in that case $\mathbf{J}$ admits a unitary eigendecomposition \cite{hu2022spectrum,preciado2016gramian,ferreira2025random}. However, for non-normal $\mathbf{J}$ like the Ginibre matrices of interest, the eigenvectors are not orthonormal, and the spectra of $\mathbf{\Sigma}_{d}$ and $\mathbf{\Sigma}_{c}$ are correspondingly harder to analyze as they are dependent on the eigenvector overlaps. While previous works have computed the asymptotic spectral distributions for related covariance matrices in other random-network or Ornstein-Uhlenbeck settings \cite{preciado2016gramian,ferreira2025random}, and considered the spectra of related covariance matrices for Ginibre $\mathbf{J}$ \cite{hu2022spectrum,shen2025covariance,tian2026diversity},\footnote{See the discussion of \citet{hu2022spectrum} in Section~\ref{sec:continuum} and Appendix~\ref{sec:hu}.} an exact characterization in the present discrete-time quenched Ginibre setting is thus far lacking. Having such a characterization would be particularly desirable in light of recent work that seeks to compare the spectra of such models to neural data \cite{pachitariu2026critical}.

The primary result of this note is a closed characterization of the spectrum of $\mathbf{\Sigma}_{d}$ in the limit $N \to \infty$. Our main result is as follows: For a stable realization of $\mathbf{J}$, let the eigenvalues of the corresponding $\mathbf{\Sigma}_{d}$ be $\lambda_{i}(\mathbf{\Sigma}_{d})$, and define their empirical distribution $\mu_{\mathbf{\Sigma}_{d}} = \frac{1}{N} \sum_{i=1}^{N} \delta_{\lambda_{i}(\mathbf{\Sigma}_{d})}$. For any fixed $0<g<1$, $\mathbf{J}$ is stable with probability tending to one as $N \to \infty$. We show that $\mu_{\mathbf{\Sigma}_{d}}$ almost surely converges weakly to a deterministic limiting spectral distribution $\mu_{d}$. Moreover, define the moments
\begin{align}
    m_{n} = \int_{0}^{\infty} \lambda^{n}\, d\mu_{d}(\lambda)
\end{align}
of the limiting spectral distribution, which almost surely give the limiting values of the moments $\frac{1}{N} \tr(\mathbf{\Sigma}_{d}^{n})$ of $\mu_{\mathbf{\Sigma}_{d}}$. We then show that the moment generating series 
\begin{equation}
    F_{d}(z) = \sum_{n=0}^{\infty} m_{n} z^{n}
\end{equation}
satisfies the scalar functional equation
\begin{equation} \label{eqn:functional_eqn}
    (1-z) F_{d}(z) = F_{d}(gz F_{d}(z))
\end{equation}
subject to the condition $F_{d}(0)=1$. 

\begin{figure}[t]
    \centering
    \includegraphics[width=3in]{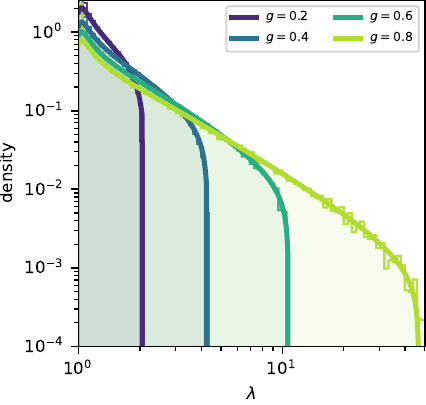}
    \caption{Comparison of the density of eigenvalues of $\mathbf{\Sigma}_d$ predicted by numerical solution of the functional equation \eqref{eqn:functional_eqn} and numerical eigendecomposition of $\mathbf{\Sigma}_d$ computed from a single realization of the random matrix $\mathbf{J}$ with $N=2048$. Solid lines show theory, and shaded stair-step patches show empirical histograms estimated by binning. Different colors show different values of the gain parameter $g$. See Appendix~\ref{sec:numerics} for details of the numerical scheme used to approximately solve \eqref{eqn:functional_eqn}.}
    \label{fig:density_comparison}
\end{figure}

Given a solution to the functional equation \eqref{eqn:functional_eqn}, we can recover the spectral distribution using the fact that the Stieltjes transform $G(s)$ of the eigenvalue density $\rho_{d}(\lambda) = d\mu_{d}/d\lambda$ is 
\begin{equation}
    G(s) = \frac{1}{s} F_d\left(\frac{1}{s}\right) ;
\end{equation}
the density follows upon Stieltjes inversion: 
\begin{equation}
    \rho_{d}(\lambda) = \lim_{\epsilon \downarrow 0} \frac{1}{\pi} \Im G(\lambda - i\epsilon) .
\end{equation}
As we will subsequently show in Section~\ref{sec:scaling}, the resulting density is compactly-supported, with a hard lower edge at $\lambda_{-}=1$, and a soft upper edge determined by the solution to an implicit equation. Solving the functional equation numerically produces an excellent match to numerical computation of the empirical spectral distribution (Figure~\ref{fig:density_comparison}). 

To our knowledge, this exact spectral characterization of $\mathbf{\Sigma}_{d}$ has not been previously reported in the literature. We derive this result using free probability techniques in Section~\ref{sec:proof}. From a random matrix theory perspective, because $\mathbf{\Sigma}_{d}$ is the solution to a Lyapunov equation with quenched randomness $\mathbf{J}$, it is not a standard Wishart- or free-convolution-type object. Nonetheless, we obtain a closed functional equation thanks to the gauge symmetry of the limiting circular element. In contrast, applying an analogous approach in the continuous-time setting yields an infinite hierarchy of equations rather than a closed scalar equation for the corresponding spectral generating function (Section~\ref{sec:continuum}). 

\section{Derivation of the discrete-time functional equation}\label{sec:proof}

A compact derivation of the claimed functional equation \eqref{eqn:functional_eqn} for $F_d$ is possible using free probability. We recall the basic ingredients required here; our treatment will be in general quite informal, and we refer the reader to the textbook of \citet{mingo2017free} or a review article by \citet{guionnet2016free} for details. Free probability deals with an algebra of random operators, equipped with a linear trace functional $\tau$ satisfying $\tau(1)=1$ with $1$ denoting the identity element, which defines expectations. The canonical example of a free random variable is a circular element. Circular elements capture the spectral properties of real matrices with i.i.d. Gaussian elements in the limit $N \to \infty$, with $\mathbf{J}$ being asymptotically replaced by $c$ and $\mathbf{J}^{\top}$ being replaced by $c^{\ast}$, and the variance condition $J_{ij} \sim \mathcal{N}(0,g/N)$ being replaced $\tau(cc^{\ast}) = g$ \cite{mingo2017free,byun2023progress}.

Given that 
\begin{align}
    \mathbf{\Sigma}_{d} = \sum_{k=0}^{\infty} \mathbf{J}^{k} (\mathbf{J}^{\top})^{k},
\end{align}
Appendix~\ref{app:convergence} establishes that its empirical spectral distribution converges as $N \to \infty$ for fixed $0<g<1$ to the spectral distribution of a free random variable $\sigma_{d}$ defined by the power series
\begin{equation}
    \sigma_{d} = \sum_{k=0}^{\infty} c^{k} (c^{\ast})^{k} . 
\end{equation}
Here, we encounter two subtleties: it is not clear whether this series is merely formal, and moreover it is not clear whether we should expect convergence of spectra, as the classical free probability results apply to non-commutative \textit{polynomials} in $\mathbf{J}$ \cite{mingo2017free}. The first of these questions can be resolved easily: this series defines a bounded operator. To see this, note that the spectral radius of the circular element $c$ is $\sqrt{g}<1$, so by Gelfand's formula for the spectral radius, $\lim_{k \to \infty} \Vert c^{k} \Vert^{1/k} = \sqrt{g}$. Thus, if we choose any $\sqrt{g} < \rho < 1$, for all sufficiently large $k$ we have $\Vert c^{k} \Vert \leq \rho^{k}$, which in turn implies that $\Vert c^{k} (c^{\ast})^{k} \Vert = \Vert c^{k} \Vert^2 \leq \rho^{2k}$, which is summable because $\rho<1$. Thus, $\sigma_{d}$ is a bounded positive operator. The question of convergence can be resolved by truncating the power series for $\mathbf{\Sigma}_{d}$ and $\sigma_{d}$ and controlling the truncation error; we defer a discussion of this point to Appendix~\ref{app:convergence}. 

We will use two important properties of circular elements in our derivation. First, circular elements have a gauge symmetry: their distribution is invariant under multiplication by complex numbers of modulus 1, \textit{i.e.}, $e^{i\phi}c$ and $c$ have the same distribution for any real $\phi$. This gauge symmetry emerges as a symmetry of the circular element. Second, circular elements admit an equivalent of Gaussian integration by parts known as ``free differencing'': for some operator $P$ we have
\begin{equation}
    \tau(c P) = g (\tau \otimes \tau) \partial_{c^{\ast}} P,
\end{equation}
where $\partial_{c^{\ast}}$ is a derivation satisfying $\partial_{c^{\ast}} c=0$, $\partial_{c^{\ast}} c^{\ast} = 1 \otimes 1$, and also the Leibniz rule
\begin{equation}
    \partial_{c^{\ast}} (UV) = (\partial_{c^{\ast}} U) (1 \otimes V) + (U \otimes 1) (\partial_{c^{\ast}} V).
\end{equation}
These facts are sufficient to derive the claimed functional equation. 

Now, as $\sigma_{d}$ is bounded and as 
\begin{equation}
    m_{n} = \tau(\sigma_{d}^{n}),
\end{equation}
the series expansion for $F_{d}(z)$ converges in an open disk $|z|<1/\Vert \sigma_{d}\Vert$, and moreover is analytic in that disk. Summing the operator-level series, we can write 
\begin{equation}
    F_{d}(z) = \tau(R(z)), \quad \textrm{where} \quad  R(z) = (1-z\sigma_{d})^{-1}.
\end{equation}
We now note that the series expansion for $\sigma_{d}$ means that it satisfies an operator-level Lyapunov equation
\begin{equation}
    \sigma_{d} = 1 + c \sigma_{d} c^{\ast}.
\end{equation}
We use this to expand $\sigma_{d} R$ in two different ways. First, the Lyapunov equation directly implies that
\begin{equation}
    \tau(\sigma_{d} R) = F_{d}(z) + \tau(c \sigma_{d} c^{\ast} R).
\end{equation}
Second, for any $z \neq 0$ we have the algebraic identity
\begin{equation}
    \tau(\sigma_{d} R) = \frac{F_{d}(z)-1}{z},
\end{equation}
which follows from the fact that 
\begin{equation}
    R-1 = (1-z\sigma_{d})^{-1} [1 - (1-z\sigma_{d})] = z \sigma_{d} R.
\end{equation}
Combining these two identities, we find that
\begin{equation}
    (1-z) F_{d}(z) - 1 = z \tau(c \sigma_{d} c^{\ast} R).
\end{equation}

Now, consider
\begin{equation}
    \nu_{n} = \tau(c \sigma_{d}^{n} c^{\ast} R),
\end{equation}
such that the term we need to compute is $\nu_{1}$. Using the integration by parts identity and the Leibniz rule, 
\begin{equation}
\begin{split}
    \nu_{n} 
    &= g (\tau\otimes\tau) [(\partial_{c^{\ast}} \sigma_{d}^{n}) (1 \otimes c^{\ast} R)] \\&\quad + g (\tau\otimes\tau) [(\sigma_{d}^{n} \otimes R)] \\&\quad + g (\tau\otimes\tau) [(\sigma_{d}^{n}c^{\ast} \otimes 1) \partial_{c^{\ast}} R].
\end{split}
\end{equation}
We consider each of these three terms in turn. First, using the formula $\sigma_{d} = 1 + c \sigma_{d} c^{\ast}$, we find that $\partial_{c^{\ast}} \sigma_{d}$ satisfies
\begin{equation}
    \partial_{c^{\ast}} \sigma_{d} = (c \otimes 1) \partial_{c^{\ast}} \sigma_{d} (1 \otimes c^{\ast}) + c \sigma_{d} \otimes 1,
\end{equation}
which upon iteration leads to the series expansion
\begin{equation}
    \partial_{c^{\ast}}\sigma_{d} = \sum_{k=0}^{\infty} c^{k+1} \sigma_{d} \otimes (c^{\ast})^{k}.
\end{equation}
Repeated application of the Leibniz rule then leads to 
\begin{equation}
    \partial_{c^{\ast}} \sigma_{d}^{n} = \sum_{k=0}^{\infty} \sum_{\ell=0}^{n-1} (\sigma_{d}^{\ell}  c^{k+1} \sigma_{d}) \otimes  ((c^{\ast})^{k}\sigma_{d}^{n-1-\ell}) .
\end{equation}
Moreover, differentiating the operator inverse yields
\begin{equation}
    \partial_{c^{\ast}} R = z (R\otimes 1) \partial_{c^{\ast}} \sigma_{d} (1 \otimes R),
\end{equation}
which gives the explicit expansion 
\begin{equation}
    \partial_{c^{\ast}} R = z \sum_{k=0}^{\infty} (R c^{k+1} \sigma_{d}) \otimes ((c^{\ast})^{k} R).
\end{equation}
Thus, the first term in $\nu_{n}$ is 
\begin{equation}
\begin{split}
    &(\tau\otimes\tau)[(\partial_{c^{\ast}} \sigma_{d}^{n}) (1 \otimes c^{\ast} R)]
    \\
    &=  \sum_{k=0}^{\infty} \sum_{\ell=0}^{n-1} \tau(\sigma_{d}^{\ell}  c^{k+1} \sigma_{d}) \tau((c^{\ast})^{k}\sigma_{d}^{n-1-\ell} c^{\ast} R)  .
\end{split}
\end{equation}
But, we note that the two traces inside the sum are not invariant under $c \mapsto e^{i\phi} c$, as they carry different numbers of powers of $c$ and $c^{\ast}$. This holds because $\sigma_{d}$ and thus $R$ carry balanced pairings of $c$ and $c^{\ast}$, hence there is nothing to counteract the extra power of $c$ in the left factor. Therefore, each trace must vanish, and thus
\begin{equation}
    (\tau\otimes\tau)[(\partial_{c^{\ast}} \sigma_{d}^{n}) (1 \otimes c^{\ast} R)] = 0.
\end{equation}
The second term is simply
\begin{equation}
    (\tau \otimes \tau)[\sigma_{d}^{n} \otimes R] = \tau(\sigma_{d}^{n}) \tau(R) = m_{n} F_{d}(z).
\end{equation}
The third term is
\begin{equation}
\begin{split}
    &(\tau \otimes \tau)[(\sigma_{d}^{n} c^{\ast} \otimes 1) \partial_{c^{\ast}} R]
    \\
    &= z \sum_{k=0}^{\infty} \tau[\sigma_{d}^{n} c^{\ast} R c^{k+1} \sigma_{d}] \tau[(c^{\ast})^{k} R].
\end{split}
\end{equation}
As we encountered in the first term, for any $k>0$ we must have $\tau[(c^{\ast})^{k} R]=0$ because $(c^{\ast})^{k}R$ is not invariant under $c \mapsto e^{i\phi} c$. This leaves us with the $k=0$ term:
\begin{align}
    (\tau \otimes \tau)[(\sigma_{d}^{n} c^{\ast} \otimes 1) \partial_{c^{\ast}} R]
    &= z \tau[\sigma_{d}^{n} c^{\ast} R c \sigma_{d}] \tau[R]
    \\
    &= z \tau[c \sigma_{d}^{n+1} c^{\ast} R] F_{d}(z)
    \\
    &= z \nu_{n+1} F_{d}(z),
\end{align}
where we have used the fact that $\tau$ is cyclic. 

Putting these results together, we conclude that 
\begin{equation}
    \nu_{n} = g m_{n} F_{d}(z) + g z \nu_{n+1} F_{d}(z).
\end{equation}
Iterating this recursion, we find that
\begin{equation}
    \nu_{1} 
    = \frac{1}{z} \sum_{n=1}^{\infty} m_{n} (g z F_{d}(z))^{n}
    = \frac{F_{d}(g z F_{d}(z))-1}{z} ,
\end{equation}
where we recognize the definition of $F_{d}$. This iteration is justified for $z$ sufficiently small; we can then extend by analyticity to all $z$ in the domain where $F_{d}$ is analytic. Therefore, 
\begin{equation}
    (1-z) F_{d}(z) - 1 = z \nu_{1} = F_{d}(gz F_{d}(z))-1,
\end{equation}
hence we conclude that, as claimed, 
\begin{equation}
    (1-z) F_{d}(z) = F_{d}(g z F_{d}(z)).
\end{equation}
We remark that the manipulations in this section can be made fully rigorous using a similar truncation argument to the one we use in Appendix~\ref{app:convergence} to show convergence of the spectra: to justify term-by-term differentiation, we can truncate the series, apply the argument as written, and then let the cutoff tend to infinity at the end, using operator norm bounds to control the remainders. 

This functional equation can also be written as a recurrence relation for the moments $m_{n}$, which shows that the condition $F_{d}(0)=1$ selects a unique formal power-series solution. To do so, we start from the expansion $F_{d}(z) = \sum_{n=0}^{\infty} m_{n} z^{n}$, where we know that $m_{0}=1$. Using this expansion, we can re-write \eqref{eqn:functional_eqn} as
\begin{equation}
    \sum_{n=1}^{\infty} (m_{n} - m_{n-1}) z^{n} = \sum_{n=1}^{\infty} \left[\sum_{k=1}^{n} m_{k} g^{k} a_{n-k}^{(k)} \right] z^{n},
\end{equation}
where we have used the $r$-fold Cauchy convolution formula to write $F_{d}(z)^{r} = \sum_{k=0}^{\infty} a_{k}^{(r)} z^{k}$ for
\begin{equation}
    a_{k}^{(r)} = \sum_{\substack{0 \leq j_{1},\ldots,j_{r} \leq k \\ j_{1} + \cdots +j_{r}=k}} m_{j_1} \cdots m_{j_r} .
\end{equation}
Equating the coefficients of $z^{n}$ for $n \geq 1$ on both sides, we find that
\begin{equation} \label{eqn:moment_recurrence}
    (1-g^{n}) m_{n} = m_{n-1} + \sum_{k=1}^{n-1} g^{k} m_{k} a_{n-k}^{(k)} .
\end{equation}
For any $0<g<1$, this is a forward recurrence that uniquely determines $m_{n}$ in terms of the lower moments $m_{0},\ldots,m_{n-1}$. This proves uniqueness of the formal power series solution, and also gives us an algorithmic procedure to compute the moments. For instance, $m_{1} = (1-g)^{-1}$, and $m_{2} = (1-g)^{-3}(1+g)^{-1}$, which allows us to compute the normalized participation ratio of the spectrum:
\begin{align}
    \lim_{N \to \infty} \frac{1}{N} \frac{\tr(\mathbf{\Sigma}_{d})^{2}}{\tr(\mathbf{\Sigma}_{d}^2)} = \frac{m_{1}^2}{m_{2}} = 1-g^2.
\end{align}
We find that the first eight moments predicted by this result agree well with numerical estimation of empirical moments (see Figure~\ref{fig:moment_comparison} in Appendix~\ref{sec:numerics}).

\section{Spectral edges and critical limits in discrete time}\label{sec:scaling}

We now turn our attention to a few properties of the limiting spectral density of $\mathbf{\Sigma}_{d}$ that we can extract from the functional equation \eqref{eqn:functional_eqn}. First, we consider the edges of the spectrum. As $\sigma_{d}$ is a bounded operator, we know that the spectrum is compactly-supported; we will show that it has a lower edge at 1 and an upper edge at some finite $\lambda_{+}$ determined by the solution to an implicit equation. Using the Stieltjes transform relationship $G(s) = F_{d}(1/s)/s$, we can re-write the functional equation as $(s-1) w = g F_{d}(w)$ upon letting $w = g  G(s)$. Formally, we can then write the solution for the inverse transform as
\begin{equation} \label{eqn:k_inverse}
    s = K(w(s)), \quad \textrm{for} \quad K(w) = 1 + \frac{g}{w}  F_{d}( w) .
\end{equation}
We can determine the spectral edges, at least implicitly, by studying the properties of this equation \cite{mingo2017free}. In particular, the spectral edges can be determined by examining when solution branches that would correspond to $s$ outside of the support of the spectrum terminate, which is made easier by the fact that $w=gG(s)$ is real for real $s$ outside of the spectral support. 

The minimum eigenvalue $\lambda_{-}$ clearly must be at least 1, as the Lyapunov equation $\sigma_{d} = 1 + c \sigma_{d} c^{\ast}$ implies that $\sigma_d \succeq 1$. The question is then whether $\lambda_{-}$ is always pinned to this lower bound. This can be directly answered in the affirmative, as the spectrum of singular values of the circular element contains zero, so as $\sigma_{d}$ is bounded and positive, the infimum of the spectrum of $c \sigma_{d} c^{\ast}$ must be zero \cite{mingo2017free}. Thus, $\lambda_{-} = 1$. 

Now we consider the maximum eigenvalue $\lambda_{+}$. For $s$ above the spectral support of $\sigma_d$, $G(s) > 0$, and moreover we have $G(s) \sim 1/s \to 0$ as $s \to \infty$. This means that the equation should have an upper branch starting from $w = 0$ (where $K(w) \to \infty$), with $w$ increasing as $s$ approaches the upper edge of the spectrum. The question is then whether $K'(w)$ can vanish for some finite $w_{+}$. Differentiating \eqref{eqn:k_inverse}, we see that this is indeed possible. Under the usual regular-edge assumption for the physical branch of the inverse Stieltjes transform \cite{mingo2017free}, the upper edge is obtained from the smallest solution $w_{+}>0$ to the equation
\begin{equation}
    w_{+} F_{d}'(w_{+}) = F_{d}(w_{+}) . 
\end{equation}
Given a solution for $w_{+}$, we then identify the upper edge of the spectrum as
\begin{equation}
    \lambda_{+} = K(w_{+}) = 1 + \frac{g}{w_{+}} F_{d}(w_{+}).
\end{equation}
Unlike the lower edge, this is a soft edge. We see in Figure~\ref{fig:density_comparison} that numerically solving for $\lambda_{+}$ produces good agreement with simulations.

We now consider the critical limit $g \uparrow 1$, where the dynamics are nearly unstable. Concretely, we suppose that we first take $N \to \infty$ at fixed $g<1$, and then take $g \uparrow 1$. Here, we can expand the functional equation to obtain an approximate solution for the long tail of large eigenvalues. Let $\delta = 1-g$, such that the critical limit is $\delta \downarrow 0$. As we expect the largest eigenvalue to diverge as $\delta \downarrow 0$, we expand $F_{d}$ in a small neighborhood around $z=0$, as the branch point at $z=1/\lambda_+$ approaches the origin. We therefore make a scaling \emph{Ansatz}
\begin{equation}
    z = \delta^{2} \zeta, \quad F_{d}(z) = 1 + \delta f(\zeta) + \delta^{2} h(\zeta) + \mathcal{O}(\delta^3) .
\end{equation}
Expanding both sides of the functional equation \eqref{eqn:functional_eqn} under this \emph{Ansatz}, we have
\begin{equation}
\begin{split}
    & 1 + \delta f(\zeta) + \delta^2 (h(\zeta)- \zeta) + \mathcal{O}(\delta^3)
    \\
    &\quad = 1 + \delta f(\zeta) + \delta^2 [h(\zeta)+f'(\zeta) (f(\zeta)-1) \zeta] + \mathcal{O}(\delta^3) . 
\end{split}
\end{equation}
The terms involving $h(\zeta)$ cancel, so we conclude that $f(\zeta)$ solves
\begin{equation}
    - 1 = f'(\zeta) (f(\zeta)-1) .
\end{equation}
This determines the leading scaling; other choices would lead to trivial scaling at leading order in $\delta$. From the fact that $F_{d}(z=0)=1$, we have the initial condition $f(\zeta=0)=0$. Then, we can solve for
\begin{equation}
    f(\zeta) = 1 - \sqrt{1-2\zeta}.
\end{equation}
This branch ceases to be real at $\zeta_{+} = 1/2$, hence we conclude that as $\delta \downarrow 0$ we have
\begin{equation}
    \lambda_{+} \sim \frac{1}{\delta^2 \zeta_{+}} = \frac{2}{\delta^2} .
\end{equation}

We therefore consider the tail density of eigenvalues on the critical scale
\begin{equation}
    \lambda = \frac{x}{\delta^2} 
\end{equation}
for $0 < x < 2$. Writing the infinitesimal quantity appearing in the Stieltjes inversion formula as $i0$, we have 
\begin{equation}
    G\left(\frac{x}{\delta^2} - i 0 \right) = \frac{\delta^2}{x} \left[1 + \delta f\left(\frac{1}{x} + i 0\right) + \mathcal{O}(\delta^2) \right].
\end{equation}
Continuing the same branch for $f(\zeta)$ into the region $\zeta > 1/2$, we see that it connects to a branch with positive imaginary part: 
\begin{equation}
    f\left(\frac{1}{x} + i 0\right) = 1 + i \sqrt{\frac{2}{x}-1} .
\end{equation}
Then, we find that
\begin{equation}
    \Im G\left(\frac{x}{\delta^2} - i 0 \right) = \frac{\delta^3}{x} \sqrt{\frac{2}{x}-1} + \mathcal{O}(\delta^4),
\end{equation}
which gives the tail density
\begin{equation}
    \frac{1}{\delta^2} \rho\left(\frac{x}{\delta^2}\right) \sim \delta \frac{\sqrt{x(2-x)}}{\pi x^2}
\end{equation}
for fixed $0<x<2$ as $\delta \downarrow 0$. Thus, near the soft upper edge of the critical tail, the tail density vanishes as $\sqrt{2-x}$. In terms of the physical eigenvalues $\lambda$, this gives a tail density
\begin{align}
    \rho(\lambda) \sim \frac{\sqrt{2}}{\pi} \lambda^{-3/2}
\end{align}
in the regime $1 \ll \lambda \ll \delta^{-2}$. 

\begin{figure*}[t]
    \centering
    \includegraphics[width=6.125in]{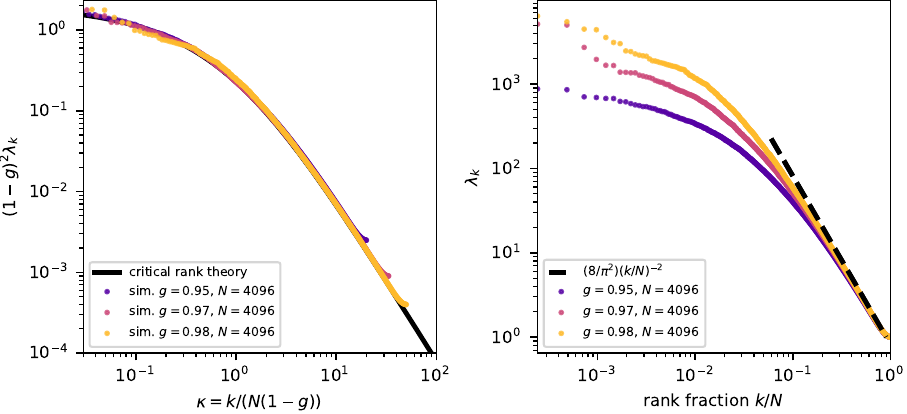}
    \caption{Scaling of eigenvalues of $\mathbf{\Sigma}_{d}$ in the critical tail. \emph{Left}: Collapse of scaled tail eigenvalues $(1-g)^2 \lambda_k$ as a function of normalized rank $\kappa = k/(N(1-g))$ across different values of $g$. \emph{Right}: In the intermediate-rank regime, the eigenvalues decay with rank approximately according to the predicted scaling law with exponent 2. As these simulations are done near the stability threshold $g=1$, we numerically exclude realizations with $\rho(\mathbf{J}) > 1-\varepsilon$ for a stability threshold $\varepsilon = 10^{-3}$. This ensures that the computation of the stationary covariance $\mathbf{\Sigma}_d$ is numerically stable.}
    \label{fig:tail_collapse}
\end{figure*}

This allows us to estimate the scaling of eigenvalues in the tail. Using the tail density, we can estimate the count of eigenvalues above some level $x/\delta^2$ by
\begin{align}
    \mathbb{P}\left[\lambda>\frac{x}{\delta^2}\right] &\sim \delta \int_{x}^{2} \frac{\sqrt{u(2-u)}}{\pi u^2} \,du \\&= \frac{2\delta}{\pi} \left[\sqrt{\frac{2-x}{x}} - \arccos \sqrt{\frac{x}{2}} \right]. \label{eqn:tail_estimate}
\end{align}
Away from the upper extreme of the tail, we can estimate the scaling by expanding near $x=0$, which gives
\begin{equation}
    \mathbb{P}\left[\lambda>\frac{x}{\delta^2}\right] \sim \delta \frac{2\sqrt{2}}{\pi \sqrt{x}} 
\end{equation}
for $\delta^2 \ll x \ll 1$. Thus, if $\lambda_k$ is the $k$-th largest among $N$ eigenvalues, for $N \delta \ll k \ll N$ we expect that
\begin{equation}
    \lambda_{k} \sim \frac{8}{\pi^2} \left(\frac{N}{k}\right)^{2}.
\end{equation}

In Figure~\ref{fig:tail_collapse}, we validate these predictions in two ways. First, we show that the ranked eigenvalues collapse almost perfectly onto the curve predicted by \eqref{eqn:tail_estimate}. In particular, if we define the normalized rank $\kappa = k/N\delta$, \eqref{eqn:tail_estimate} is of the form $\kappa = R(x)$, which gives us a theoretical prediction for the eigenvalue as a function of rank. Second, we show that the simple inverse-squared scaling of eigenvalue with normalized rank gives good predictions for the scaling in an intermediate regime, with corrections visible for the largest and smallest eigenvalues. This is consistent with the expectation that the scaling should hold for $N\delta \ll k \ll N$.

\section{Continuous-time dynamics and the loss of scalar closure}\label{sec:continuum}

Now, we show why it is harder to analyze the limiting spectrum of $\mathbf{\Sigma}_c$. What we will find is that the analogous free-differencing calculation leads to an infinite hierarchy of Schwinger-Dyson equations, rather than a closed scalar equation for the generating function itself. This difference reflects the fact that $\sigma_{d}$ is gauge-neutral as a function of $c$ and $c^{\ast}$---each term is invariant under $c \mapsto e^{i\theta} c$---while $\sigma_{c}$ is not. 

Before embarking on the analysis, we note that the stationary covariance $\mathbf{\Sigma}_{c}$ considered here is distinct from the frequency-domain covariance $\mathbf{Q}(\omega)$ studied by \citet{hu2022spectrum}. This object is obtained by taking the Fourier transform of the stationary time-lagged covariance $\mathbf{Q}(\tau) = \lim_{t \to \infty} \mathbb{E}[x(t)x(t+\tau)^{\top}]$, and is given for a linear network by
\begin{equation}
    \mathbf{Q}(\omega) = [(1+i\omega) \mathbf{I}_{N} - \mathbf{J}]^{-1}  [(1-i\omega) \mathbf{I}_{N} - \mathbf{J}^{\top}]^{-1} .
\end{equation}
To relate $\mathbf{Q}(\omega)$ to $\mathbf{\Sigma}_{c}$, we apply Plancherel's theorem to the matrix-valued function $\Theta(t) e^{-(I-\mathbf{J})t}$ \cite{gardiner1985handbook}, which gives
\begin{equation} \label{eqn:sigma_c_fourier}
    \mathbf{\Sigma}_{c} = \int_{-\infty}^{\infty} \mathbf{Q}(\omega) \frac{d\omega}{2\pi} .
\end{equation}
As $\mathbf{J}$ is non-normal, the eigenvectors of $\mathbf{Q}(\omega)$ are $\omega$-dependent, meaning that the spectral density of $\mathbf{Q}(\omega)$ does not directly determine that of $\mathbf{\Sigma}_{c}$ \cite{chalker1998eigenvector}. Thus, while \citet{hu2022spectrum} obtained an interpretable expression for the spectral density of $\mathbf{Q}(\omega)$, we do not obtain a similarly-compact characterization of the spectrum of $\mathbf{\Sigma}_c$. We will return to this distinction in Appendix~\ref{sec:hu}, where we give a compact derivation of the Hu-Sompolinsky density.

Returning to $\mathbf{\Sigma}_c$, the relevant free random variable is
\begin{equation}
    \sigma_{c} = \int_{0}^{\infty} e^{c t} e^{c^{\ast} t} e^{-2t}\, dt,
\end{equation}
which satisfies
\begin{equation}
    \sigma_{c} = \frac{1}{2} + \frac{1}{2} (c \sigma_{c} + \sigma_{c} c^{\ast}). 
\end{equation}
As the spectral radius of $c$ is bounded, the integral converges to a bounded operator. Using an analogous approach to the one we used in discrete time, we can show that the spectrum of $\mathbf{\Sigma}_{c}$ converges almost surely (both weakly and in the sense of moments) to that of $\sigma_{c}$ for any fixed $0<g<1$ (Appendix~\ref{app:convergence}). This shows that the challenge in continuous time is not identifying the relevant free operator, but rather obtaining a simple description of the spectrum of said operator. 

By expanding the two exponentials and evaluating the integral over time, we obtain a series expansion 
\begin{equation} \label{eqn:sigma_c_series}
    \sigma_{c} = \sum_{j,k=0}^{\infty} \alpha_{jk} c^{j} (c^{\ast})^{k},
\end{equation}
where
\begin{equation} \label{eqn:alpha_matrix}
    \alpha_{jk} = \frac{1}{j!\,k!} \int_{0}^{\infty} t^{j+k} e^{-2t}\,dt = \frac{1}{2^{j+k+1}} \binom{j+k}{k} .
\end{equation}
As in the series expansion for $\sigma_{d}$, this series defines a bounded operator thanks to the fact that, for any $\sqrt{g} < \rho < 1$, for all $j$ and $k$ large enough we have $\Vert c^{j} (c^{\ast})^{k} \Vert \leq \rho^{j+k}$. However, because it contains terms with different powers of $c$ and $c^{\ast}$, it is not invariant under $c \mapsto e^{i\phi} c$. This will turn out to be the source of the obstruction to obtaining scalar closure as we did in the discrete case. $\sigma_{d}$ (and thus its resolvent) are neutral under gauge transformations $c \mapsto e^{i \phi }c$, which means that traces of powers of $c$ against the resolvent vanish by gauge symmetry. In contrast, $\sigma_{c}$ is not gauge-neutral, so the analogous correlators are not forced to vanish by the gauge symmetry. 

Now, we proceed with the analysis. Analogously to the discrete-time case, we let 
\begin{equation}
    F_{c}(z) = \tau(R(z)), \quad \textrm{where} \quad  R(z) = (1-z\sigma_{c})^{-1}.
\end{equation}
Again, for any $z \neq 0$ we have the algebraic identity
\begin{equation}
    \tau(\sigma_{c} R) = \frac{F_{c}(z)-1}{z}  ; 
\end{equation}
what is different is that we now obtain the expansion 
\begin{equation}
    \tau(\sigma_{c} R) 
    = \frac{1}{2} F_{c}(z) + \frac{1}{2} [\tau(c \sigma_{c} R) + \tau(\sigma_{c} c^{\ast} R)] 
\end{equation}
from the Lyapunov equation, where in the second line we have used the fact that $R$ and $\sigma_{c}$ commute. This leads to the equation
\begin{equation}
    F_{c}(z) - 1 - \frac{1}{2} z F_{c}(z) = \frac{1}{2} z [\tau(c \sigma_{c} R) + \tau(\sigma_{c} c^{\ast} R)]  
\end{equation}
which, using the algebraic identity $z \sigma_{c} R = R-1$ and the fact that $\tau(c) = \tau(c^{\ast}) = 0$, simplifies to
\begin{equation}
    F_{c}(z) - 1 - \frac{1}{2} z F_{c}(z) = \frac{1}{2}  [\tau(c R)  + \tau(c R(z^{\ast}))^{\ast}] ,
\end{equation}
where we use the fact that $\tau(c^{\ast} R) = \tau(c R(z^{\ast}))^{\ast}$.

We now integrate by parts: 
\begin{align}
    \tau(c R) &= g (\tau \otimes \tau) \partial_{c^{\ast}} R
    \\
    &= g z (\tau \otimes \tau) [(R\otimes 1) \partial_{c^{\ast}} \sigma_{c} (1 \otimes R)].
\end{align}
To determine $\partial_{c^{\ast}}\sigma_{c}$, we can differentiate the Lyapunov equation to get
\begin{equation}
    \partial_{c^{\ast}}\sigma_{c} = \frac{1}{2} (\sigma_{c} \otimes 1) + \frac{1}{2} (c \otimes 1) \partial_{c^{\ast}} \sigma_{c} + \frac{1}{2} \partial_{c^{\ast}} \sigma_{c} (1 \otimes c^{\ast}),
\end{equation}
which on iteration leads to the series expansion
\begin{equation}
    \partial_{c^{\ast}} \sigma_{c} = \sum_{j,k=0}^{\infty} \alpha_{jk} (c^{j} \sigma_{c}) \otimes (c^{\ast})^{k} ,
\end{equation}
where $\alpha_{jk}$ is as in \eqref{eqn:alpha_matrix}. Substituting this expansion into the formula for $\tau(c R)$, we have
\begin{equation}
    \tau(c R) = g z  \sum_{j,k=0}^{\infty} \alpha_{jk} \tau(R c^{j} \sigma_{c}) \tau( (c^{\ast})^{k} R ).
\end{equation}
However, unlike in the discrete case, we cannot use distributional invariance under $c \mapsto e^{i\phi} c$ to argue that all but a few of the terms in this sum vanish, because $\sigma_{c}$ itself contains both gauge-invariant and gauge-variant terms. This is evident from the series expansion \eqref{eqn:sigma_c_series}.

Therefore, within this free-differencing approach, we do not obtain a simple closed equation for $F_{c}(z)$. Instead, we find that it is coupled to an infinite family of correlators like $\tau(c^{k} R)$ and $\tau(R c^{j} \sigma_{c})$. We could in principle write down an infinite set of coupled Schwinger-Dyson equations that together determine these objects---as the equations relating traces of these correlators are of Schwinger-Dyson type \cite{guionnet2016free}. This would allow us to algorithmically compute statistics like moments of the spectral distribution---with the complexity of the computation increasing rapidly as we go to higher moments---but the simple scalar reduction is lost. We will therefore content ourselves with the fact that we have a clear picture for why the equation for the generating function does not close in the same simple way as it did in the discrete case.

\section{Discussion}

We have shown that the stationary covariance spectrum of a discrete-time linear recurrent neural network with non-normal random weights satisfies a simple functional equation \eqref{eqn:functional_eqn}. This allows us to characterize the distribution of variance among principal components, and extract the long tail of eigenvalues in the critical regime $g \uparrow 1$ (Section~\ref{sec:scaling}). In contrast, we found that applying the same free probability approach to the analogous continuous-time dynamics leads to an infinite hierarchy of Schwinger-Dyson equations, rather than a single scalar functional equation (Section~\ref{sec:continuum}). We now comment on some of the implications of our results, and potential directions for future inquiry. 

As we noted in Section~\ref{sec:continuum}, the distinction in complexity of the characterization of the spectral distribution in discrete and continuous time arises from a difference in the symmetries of the stationary covariances. $\mathbf{\Sigma}_d$ is a sum of balanced terms $\mathbf{J}^{k} (\mathbf{J}^{\top})^{k}$, which translate into gauge-invariant terms $c^{k} (c^{\ast})^{k}$ at the free probability level. In contrast, $\mathbf{\Sigma}_c$ includes all balanced and imbalanced terms $\mathbf{J}^{j} (\mathbf{J}^{\top})^{k}$, which leads to gauge-dependent terms $c^{j} (c^{\ast})^{k}$ for $j \neq k$. The gauge-invariance in the discrete case enables us to show that all but a finite number of terms in the expansions we encounter vanish, leading to a closed equation for the generating function $F_d$, while no such reduction to a closed equation for $F_c$ is immediately apparent in the continuum. This obstruction is consistent with the broader role of eigenvector non-orthogonality in non-normal Langevin dynamics, where stationary correlations, response functions, and entropy production depend explicitly on Chalker-Mehlig-type eigenvector overlaps \cite{chalker1998eigenvector,fyodorov2025nonorthogonal,hu2022spectrum,marti2018correlations}. As our main focus is on the simplicity of the discrete-time result, we have not endeavored to perform a further detailed analysis of the continuous-time hierarchy. This will be an important and interesting topic for future work. 

The contrast between the discrete- and continuous-time dynamics can be sharpened by considering the Euler-Maruyama discretization of the continuous dynamics. Discretizing \eqref{eqn:continuous} with a step size $\eta>0$ leads to 
\begin{align}
    \mathbf{x}_{t+1} = [(1-\eta) \mathbf{I}_{N} + \eta \mathbf{J}] \mathbf{x}_{t} + \sqrt{\eta} \mathbf{z}_{t}
\end{align}
so the stationary covariance $\mathbf{\Sigma}_{\eta}$ of the discretized process satisfies the Lyapunov equation
\begin{equation}
    \mathbf{\Sigma}_{\eta} = \eta \mathbf{I}_{N} + [(1-\eta) \mathbf{I}_{N} + \eta \mathbf{J}] \mathbf{\Sigma}_{\eta} [(1-\eta) \mathbf{I}_{N} + \eta \mathbf{J}]^{\top}. 
\end{equation}
The intrinsic discrete-time dynamics \eqref{eqn:discrete} is recovered by taking $\eta = 1$. However, for any $0< \eta < 1$, the $(1-\eta) \mathbf{I}_{N}$ terms will generate terms in the solution for $\mathbf{\Sigma}_{\eta}$ that involve mixed powers $\mathbf{J}^{j}(\mathbf{J}^{\top})^{k}$ with $j \neq k$. Thus, the gauge symmetry at $\eta = 1$ is lost for any $\eta < 1$. It is therefore the particular discrete-time update rule \eqref{eqn:discrete}, rather than discreteness itself, that enables scalar closure. 

The Lyapunov equation $\sigma_d = 1 + c \sigma_{d} c^{\ast}$ resembles a class of fixed-point equations known as matrix or free \emph{perpetuities}, which have very recently been studied in the random matrix literature \cite{belinschi2025free,kolodziejek2026empirical}. These equations take the form of distributional affine fixed-point equations $\mathbf{X} \overset{d}{=} \mathbf{A} + \mathbf{B} \mathbf{X} \mathbf{B}^{\top}$, where $(\mathbf{A},\mathbf{B})$ is a pair of random matrices or free random variables. However, the solution $\mathbf{X}$ to the perpetuity is assumed to be independent (or, in the limit, free) of the pair $(\mathbf{A},\mathbf{B})$. This is manifestly not true of the Lyapunov equation studied here, as $\sigma_{d}$ is defined by the solution to the Lyapunov equation for a fixed realization of the circular element $c$. In other words, the randomness in the Lyapunov equation is quenched, while in some sense in a perpetuity it is annealed. With this distinction in mind, it would be interesting to connect the spectral laws for free perpetuities to those for the solutions to random Lyapunov equations. 

More broadly, random matrix theory has long concerned itself with the development of models for the spectra of observed covariance matrices \cite{potters2020first,marchenko1967distribution,silverstein1995analysis}. Classical results going back to the Marchenko-Pastur theorem characterize how the spectrum of a population covariance matrix is distorted by sampling a limited number of observations \cite{marchenko1967distribution,silverstein1995analysis}. With the advent of large-scale recordings of neural population activity, this question of covariance estimation under observational limits has attracted significant attention in neuroscience \cite{hu2022spectrum,landau2026predictive,gao2017theory}. This literature---both in the neuroscience context and beyond---is extensive, and a complete review would be beyond the scope of the present work. We emphasize, however, that it is largely focused on the question of how a given population covariance spectrum is observed \cite{marchenko1967distribution,silverstein1995analysis,landau2026predictive} or phenomenologically modeled \cite{potters2020first,levi2025statistical}. Here, we ask a complementary bottom-up question of how underlying microscopic dynamics generate a particular population covariance spectrum. 

With this broader literature in mind, we now return to the neuroscientific motivation of our study: the use of stationary, random linear recurrent dynamics as a null model for principal component spectra in richer recurrent models and in recordings of neural activity. In this regard, our work is substantially motivated by a recent paper of \citet{pachitariu2026critical}, which compared the stationary covariance spectrum of continuous-time dynamics with varying choices of connectivity matrix ensemble to measurements of brainwide activity in mice. Based on simulations, \citet{pachitariu2026critical} conjectured that the critical tail of eigenvalues of $\mathbf{\Sigma}_{c}$ decays with rank with an exponent of $5/4$. A very recent paper by \citet{feng2026near} has used a direct asymptotic study of the moments of the continuous-time law in the critical limit $g \uparrow 1$ to establish that the rank-scaling exponent is 4/3, not 5/4 as conjectured by \citet{pachitariu2026critical}. Our results here show that the analogous discrete-time model has a yet larger ranked-eigenvalue exponent of $2$. This is in further tension with their finding that neural recordings across several cortical and sub-cortical areas show decay exponents between $0.75$ and $0.8$, which are closer to, but do not coincide with, the $2/3$ exponent predicted by taking $\mathbf{J}$ to be a \emph{symmetric} Gaussian random matrix. Our results thus underscore the fact that, even in null models, one can substantially change the statistics of critical dynamics through seemingly-innocuous adjustments, including the choice of dynamical update rule \cite{chen2024searching}. 

Finally, we note in closing that we have restricted our attention to linear recurrent dynamics. This enabled us to leverage the exact finite-$N$ expressions available for the stationary covariance in the linear setting, which in turn allowed the direct application of free probability. However, nonlinear random recurrent neural networks would of course constitute a much richer class of null models for principal component analysis of stationary activity, as they can display true chaotic behavior \cite{sompolinsky1988chaos}. In a pair of recent works, \citet{clark2026linear} and \citet{wakhloo2026solution} have shown that the long-time-window covariance of a nonlinear network with random weights can, in the $N \to \infty$ limit, be replaced by an effective linear network. This result proves a conjecture by \citet{shen2025covariance}, who used a linearization of this form to extend the \citet{hu2022spectrum} spectral density from linear to nonlinear networks. These equivalences are analogous to an ever-expanding array of so-called Gaussian equivalence theorems that have been proved for random feedforward neural networks \cite{lu2022equivalence,atanasov2026scaling}. In light of the results we have obtained here, it would be interesting to extend these equivalences and the resulting spectral characterizations to the stationary setting.

\section*{Acknowledgements}

The author thanks Blake Bordelon, David Clark, and Noam Levi for inspiring discussions and helpful comments. This work was supported by the Office of the Director of the NIH under Award Number DP5OD037354, and by a Junior Fellowship from the Harvard Society of Fellows.

\section*{AI Usage Declaration}

OpenAI models GPT 5.5 Pro, 5.6 Sol, and 6 Sol assisted in the preparation of this manuscript. GPT 5.5 Pro was able to independently derive the central functional equation, prompted with only the suggestion to apply integration by parts to the discrete-time model and without access to the author's initial derivation sketch. The author used the AI models to generate the initial version of the code used to make Figures \ref{fig:density_comparison} and \ref{fig:tail_collapse}, to suggest improvements in the formulation and presentation of the results, and to check the manuscript for mathematical and typographical errors. The models correctly pointed out several gaps of rigor, and helped simplify and clarify the derivations. GPT 6 Sol was particularly helpful in improving the convergence argument in Appendix~\ref{app:convergence}. The final arguments and presentation are the author's own, and he takes responsibility for the content. 

\section*{Data and code availability}

Code to reproduce all figures is available on GitHub at \url{https://github.com/jzavatoneveth/discrete-nonnormal-dynamics-spectrum}. No external datasets were used; the figures depend only on the results of numerical simulations implemented in the code on GitHub. 

\appendix 

\section{Spectral convergence}\label{app:convergence}

Here, we return to the question of convergence of the spectrum of $\mathbf{\Sigma}_d$ to that of $\sigma_d$ for typical realizations at fixed $g<1$. Recalling that $\mathbf{\Sigma}_{d}$ and $\sigma_{d}$ are defined by infinite series, this is not necessarily obvious, because the standard free probability result is that non-commutative \textit{polynomials} in $\mathbf{J}$ converge to the corresponding polynomials in $c$ as $N \to \infty$ \cite{mingo2017free}. However, we can use an argument based on truncation of the power series to arrive at the desired conclusion. We then finally show that a similar argument can be applied to show convergence of the spectrum of $\mathbf{\Sigma}_{c}$ to that of $\sigma_{c}$. 

\subsection{A preliminary lemma: operator norm bounds}

As a preliminary step, we prove a useful lemma that will allow us to control operator norms of powers of $\mathbf{J}$ and of $c$. The key idea is that submultiplicative property of the operator norm allows us to convert bounds that hold for any fixed power of either operator into those that hold for all powers simultaneously. Consider a general sequence of matrices $A_{N}$, with dimensions possibly depending on $N$, such that 
\begin{align}
    \Vert A_{N}^{k} \Vert \to \Vert a^{k} \Vert
\end{align}
as $N \to \infty$ for each fixed integer $k \geq 1$, where $a$ is a bounded operator. If the spectral radius $r(a)$ of $a$ satisfies $r(a)<1$, then what we will show is that there are constants $C < \infty$, $0<\rho<1$, and an integer $N_{0}<\infty$ such that 
\begin{align}
    \max(\Vert A_{N}^{k} \Vert, \Vert a^{k} \Vert) \leq C \rho^{k}
\end{align}
for all $N \geq N_{0}$ and $k \geq 0$. Importantly, this bound holds simultaneously for all $k$, not just for any fixed $k$. 

To prove this, we start by noting that Gelfand's formula for the spectral radius, $\lim_{k \to \infty} \Vert a^{k} \Vert^{1/k} = r(a)$, means that we may choose a fixed integer $K$ such that $\Vert a^{K} \Vert < 1$. Given this fixed $K$, we can choose a $0<q<1$ such that $\Vert a^{K} \Vert < q$. By the assumption that $\Vert A_{N}^{k} \Vert \to \Vert a^{k} \Vert$, there thus must exist some $N_{0}$ such that $\Vert A_{N}^{K} \Vert \leq q$ for all $N \geq N_{0}$. Moreover, as $K$ is fixed, we know that all of the norms up to $K$ must converge by assumption, and therefore given the $N_{0}$ chosen above there exists a constant $B < \infty$ such that
\begin{align}
    \max_{0 \leq r < K}\{ \Vert A_{N}^{r} \Vert, \Vert a^{r} \Vert \} \leq B
\end{align}
for all $N \geq N_{0}$. 

Now, for an arbitrary integer $k \geq 0$, write $k = m K + r$, where $0 \leq r < K$ and $m = \lfloor k/K \rfloor$. By the submultiplicative property of the operator norm, 
\begin{align}
    \Vert A_{N}^{k} \Vert \leq \Vert A_{N}^{K} \Vert^{m} \Vert A_{N}^{r} \Vert \leq B q^{m},
\end{align}
where the final inequality uses those obtained above and thus applies for all $N \geq N_{0}$. An identical argument shows that $\Vert a^{k} \Vert\leq B q^{m}$. Then, as $m = \lfloor k/K \rfloor \geq k/K - 1$, we know that $B q^{m} \leq B q^{-1} (q^{1/K})^{k} = C \rho^{k}$, where we let $C=Bq^{-1} < \infty$ and $\rho = q^{1/K} < 1$. Thus, we find that, as desired
\begin{align}
    \max(\Vert A_{N}^{k} \Vert, \Vert a^{k} \Vert) \leq C \rho^{k}. 
\end{align}
Here, this bound holds for all $k$: we have used the submultiplicative property of the operator norm to convert a bound obtained for the fixed order $K$ and each order below it into one that holds simultaneously for all $k$.

\subsection{Convergence of spectra in discrete time}

With the lemma above in hand, we are ready to prove the desired convergence statement. Fix $0<g<1$, and recall the series definitions
\begin{align}
    \mathbf{\Sigma}_{d} &= \sum_{k=0}^{\infty} \mathbf{J}^{k} (\mathbf{J}^{\top})^{k}, 
    \\
    \sigma_{d} &= \sum_{k=0}^{\infty} c^{k} (c^{\ast})^{k}. 
\end{align}
Then, for all sufficiently large $N$, the infinite series defining $\mathbf{\Sigma}_{d}$ converges to a bounded operator almost surely. Moreover, we have that for any continuous test function $\varphi$ 
\begin{align}
    \frac{1}{N} \tr \varphi( \mathbf{\Sigma}_{d} ) \to \tau( \varphi(\sigma_{d} ) )
\end{align}
and that
\begin{align}
    \Vert \varphi(\mathbf{\Sigma}_{d}) \Vert \to \Vert \varphi(\sigma_{d}) \Vert ,
\end{align}
where both statements hold almost surely as $N \to \infty$. This proves almost-sure convergence of the empirical spectral distributions, of the moments, and also convergence of the endpoints of their support. To be very explicit, setting $\varphi(x) = x$ allows us to conclude convergence of the maximum eigenvalue, and for the minimum eigenvalue we may take $\varphi(x) = M-x$ for $M$ a common upper bound on the spectral support. 

To prove this claim, our starting point is the result of \citet{schultz2005non}: for any non-commutative polynomial $P$, we have almost-sure convergence of traces and operator norms. In other words, as $N \to \infty$ we have 
\begin{align}
    \frac{1}{N} \tr P(\mathbf{J}, \mathbf{J}^{\top}) &\to \tau( P(c,c^{\ast}) ),
    \\
    \Vert P(\mathbf{J}, \mathbf{J}^{\top}) \Vert &\to \Vert P(c,c^{\ast}) \Vert
\end{align}
almost surely. Henceforth, we condition on the probability-one event on which this strong convergence holds for all non-commutative polynomials, as guaranteed by \cite{schultz2005non}.  

As noted in the main text, the spectral radius of the circular element of variance $g$ is $\sqrt{g} < 1$. Thus, using \citet{schultz2005non}'s result in the special case of the monomials $\mathbf{J}^{k}$ and $c^{k}$, we can immediately invoke the lemma proved in the prelude to conclude that there exist $C<\infty$, $0<\rho<1$, and $N_{0} < \infty$ such that
\begin{align}
    \Vert \mathbf{J}^{k} \Vert , \Vert c^{k} \Vert \leq C \rho^{k}
\end{align}
for all $N \geq N_{0}$ and all $k \geq 0$. Now, for $L \geq 0$, define the truncated series
\begin{align}
    \mathbf{\Sigma}_{L} = \sum_{k=0}^{L} \mathbf{J}^{k} (\mathbf{J}^{\top})^{k}, \quad \sigma_{L} = \sum_{k=0}^{L} c^{k} (c^{\ast})^{k}. 
\end{align}
Using the bound above, we have that $\Vert \mathbf{J}^{k} (\mathbf{J}^{\top})^{k} \Vert = \Vert \mathbf{J}^{k} \Vert^2 \leq C^{2} \rho^{2k}$. Therefore, using the triangle inequality and summing the resulting geometric series, 
\begin{align}
    \Vert \mathbf{\Sigma}_{d} - \mathbf{\Sigma}_{L} \Vert = \left\Vert \sum_{k=L+1}^{\infty} \mathbf{J}^{k} (\mathbf{J}^{\top})^{k} \right\Vert \leq \frac{C^2 \rho^{2(L+1)}}{1-\rho^2}
\end{align}
and similarly
\begin{align}
    \Vert \sigma_{d} - \sigma_{L} \Vert \leq \frac{C^2 \rho^{2(L+1)}}{1-\rho^2}. 
\end{align}
An analogous argument shows also that
\begin{align}
    \Vert \mathbf{\Sigma}_{d} \Vert, \Vert \sigma_{d}\Vert, \Vert \mathbf{\Sigma}_{L} \Vert, \Vert \sigma_{L} \Vert \leq \frac{C^2}{1-\rho^2}. 
\end{align}
Together, these bounds establish stability of $\mathbf{J}$ for all sufficiently large $N$, and convergence in operator norm of the series defining $\mathbf{\Sigma}_{d}$ and $\sigma_{d}$. Importantly, we have shown that the truncation error tends to zero as $L \to \infty$ uniformly in $N \geq N_{0}$. 

We can now prove convergence for polynomial observables $p$. By the triangle inequality and the fact that normalized traces are bounded by operator norms, 
\begin{align}
    &\left| \frac{1}{N} \tr p(\mathbf{\Sigma}_{d}) - \tau( p( \sigma_{d} ) ) \right | 
    \nonumber\\&\leq\left| \frac{1}{N} \tr p(\mathbf{\Sigma}_{L}) - \tau( p( \sigma_{L} ) ) \right | \nonumber\\&\quad + \Vert p(\mathbf{\Sigma}_{d}) - p(\mathbf{\Sigma}_{L}) \Vert + \Vert p(\sigma_{d})  - p(\sigma_{L}) \Vert. 
\end{align}
Similarly, using the reverse triangle inequality for operator norms,
\begin{align}
    &| \Vert p(\mathbf{\Sigma}_{d}) \Vert - \Vert p( \sigma_{d} ) \Vert  | 
    \nonumber\\&\leq | \Vert p(\mathbf{\Sigma}_{L}) \Vert - \Vert p( \sigma_{L} ) \Vert | \nonumber\\&\quad + \Vert p(\mathbf{\Sigma}_{d}) - p(\mathbf{\Sigma}_{L}) \Vert + \Vert p(\sigma_{d})  - p(\sigma_{L}) \Vert. 
\end{align}
For any fixed $L$, $\mathbf{\Sigma}_{L}$ is a non-commutative polynomial in $\mathbf{J}$ and $\mathbf{J}^{\top}$, so \citet{schultz2005non}'s result directly gives
\begin{align}
    \frac{1}{N} \tr p(\mathbf{\Sigma}_{L}) &\to \tau( p(\sigma_{L}) ), \\
    \Vert p(\mathbf{\Sigma}_{L}) \Vert &\to \Vert p(\sigma_{L}) \Vert
\end{align}
as $N \to \infty$. 

What remains is to control the error terms $\Vert p(\mathbf{\Sigma}_{d}) - p(\mathbf{\Sigma}_{L}) \Vert$ and $\Vert p(\sigma_{d})  - p(\sigma_{L}) \Vert$. First consider $\Vert p(\mathbf{\Sigma}_{d}) - p(\mathbf{\Sigma}_{L}) \Vert$; the argument for $\Vert p(\sigma_{d})  - p(\sigma_{L}) \Vert$ is analogous. For a generic degree-$q$ polynomial $p(x) = \sum_{j=0}^{q} \alpha_{j} x^{j}$ with real coefficients $\alpha_{j}$, we can use the triangle inequality and the standard telescoping bound for the norm of a difference of powers to obtain 
\begin{align}
    &\Vert p(\mathbf{\Sigma}_{d}) - p(\mathbf{\Sigma}_{L}) \Vert \nonumber\\ &\leq \sum_{j=1}^{q} |\alpha_{j}| \Vert \mathbf{\Sigma}_{d}^{j} - \mathbf{\Sigma}_{L}^{j} \Vert 
    \\
    &\leq \sum_{j=1}^{q} |\alpha_{j}| \sum_{\ell=0}^{j-1} \Vert \mathbf{\Sigma}_{d} \Vert^{j-1-\ell} \Vert \mathbf{\Sigma}_{L} \Vert^{\ell} \Vert \mathbf{\Sigma}_{d} - \mathbf{\Sigma}_{L} \Vert 
    \\
    &\leq \left[ \sum_{j=1}^{q} |\alpha_{j}| j \left(\frac{C^2}{1-\rho^2} \right)^{j-1} \right] \frac{C^2 \rho^{2(L+1)}}{1-\rho^2}, 
\end{align}
as we know that $\Vert \mathbf{\Sigma}_{d} \Vert, \Vert \mathbf{\Sigma}_{L} \Vert \leq \frac{C^2}{1-\rho^2}$ and that $\Vert \mathbf{\Sigma}_{d} - \mathbf{\Sigma}_{L} \Vert \leq \frac{C^2 \rho^{2(L+1)}}{1-\rho^2}$. An analogous argument shows that an identical upper bound holds for $\Vert p(\sigma_{d})  - p(\sigma_{L}) \Vert$. These bounds hold for all $N \geq N_0$, hence we may first take $N \to \infty$ and then $L \to \infty$ to conclude that the two error terms tend to zero. 

Therefore, we conclude that for any polynomial $p$ we have 
\begin{align}
    \frac{1}{N} \tr p( \mathbf{\Sigma}_{d} ) &\to \tau( p(\sigma_{d} ) )
    \\
    \Vert p(\mathbf{\Sigma}_{d}) \Vert &\to \Vert p(\sigma_{d}) \Vert ,
\end{align}
almost surely as $N \to \infty$. Now, from the operator norm bounds before, we know that the spectral support of all operators of interest ($\mathbf{\Sigma}_{d}$, $\sigma_{d}$, $\mathbf{\Sigma}_{L}$, and $\sigma_{L}$) must be contained in the fixed compact interval $[0, \frac{C^2}{1-\rho^2}]$. Therefore, by the Stone–Weierstrass Theorem for uniform approximation of continuous functions over compact intervals, we can extend these results to any continuous function $\varphi$, thus proving the claim.

\subsection{Convergence of spectra in continuous time}

The same truncation strategy that allowed us to show almost-sure convergence of the spectrum of $\mathbf{\Sigma}_{d}$ to that of $\sigma_{d}$ extends quite directly to a proof of convergence of the spectrum of $\mathbf{\Sigma}_{c}$ to that of $\sigma_{c}$. Recall that 
\begin{align}
    \mathbf{\Sigma}_{c} &= \int_{0}^{\infty} e^{-2t} e^{\mathbf{J} t} e^{\mathbf{J}^{\top} t}\, dt, 
    \\
    \sigma_{c} &= \int_{0}^{\infty} e^{-2t} e^{c t} e^{c^{\ast} t} \, dt. 
\end{align}
As shown before, there exist $C<\infty$, $0<\rho<1$, and $N_{0} < \infty$ such that
\begin{align}
    \Vert \mathbf{J}^{k} \Vert , \Vert c^{k} \Vert \leq C \rho^{k}
\end{align}
almost surely, simultaneously for all $k$. Thus, 
\begin{align}
    \Vert e^{t \mathbf{J}} \Vert \leq \sum_{k=0}^{\infty} \frac{t^{k}}{k!} \Vert \mathbf{J}^{k} \Vert \leq C e^{\rho t}
\end{align}
and similarly $\Vert e^{ct} \Vert \leq C e^{\rho t}$. As a result, 
\begin{align}
    \Vert  e^{-2t} e^{\mathbf{J} t} e^{\mathbf{J}^{\top} t} \Vert, \ \Vert e^{-2t} e^{c t} e^{c^{\ast} t} \Vert \leq C^{2} e^{-2 (1-\rho) t} . 
\end{align}
As $\rho<1$, this bound decays exponentially in time. Thus, if we fix a truncation time $T>0$, we find that
\begin{align}
    \left\Vert \mathbf{\Sigma}_{c} - \int_{0}^{T} e^{-2t} e^{\mathbf{J} t} e^{\mathbf{J}^{\top} t}\,dt\right\Vert \leq \frac{C^2 e^{-2(1-\rho)T}}{2(1-\rho)}
\end{align}
uniformly in $N \geq N_{0}$, and similarly for the integral defining $\sigma_{c}$. Using the fact that $\Vert \mathbf{J} \Vert$ and $\Vert c \Vert$ are uniformly bounded, we can uniformly approximate the integral over $[0,T]$ by a non-commutative Taylor polynomial of fixed order $M$. This at last gives us a polynomial to which we can apply the convergence theorem of \citet{schultz2005non}. So, taking $N \to \infty$, then $M \to \infty$, and finally $T \to \infty$, we conclude that
\begin{align}
    \frac{1}{N} \tr \varphi(\mathbf{\Sigma}_{c}) &\to \tau(\varphi(\sigma_{c})),
    \\
    \Vert \varphi(\mathbf{\Sigma}_{c}) \Vert &\to \Vert \varphi(\sigma_{c}) \Vert,
\end{align}
again relying on Stone-Weierstrass to pass from polynomial to continuous test functions.

\section{A tractable question in continuous time: recovering the Hu-Sompolinsky density}\label{sec:hu}

As a counterpoint to the lack of scalar closure we found in Section~\ref{sec:continuum} for $\mathbf{\Sigma}_c$, we now show that the free probability approach allows us to give a very compact derivation of \citet{hu2022spectrum}'s result for the spectrum of the frequency-domain covariance
\begin{equation}
    \mathbf{Q}(\omega) = [(1+i\omega) \mathbf{I}_{N} - \mathbf{J}]^{-1}  [(1-i\omega) \mathbf{I}_{N} - \mathbf{J}^{\top}]^{-1} .
\end{equation}
For a non-zero complex number $\zeta$ such that $|\zeta|>\sqrt{g}$, define the operator 
\begin{equation}
    q(\zeta) = (\zeta - c)^{-1} (\zeta^{\ast} - c^{\ast})^{-1}. 
\end{equation}
Then, as $N \to \infty$, the empirical spectral distribution of $\mathbf{Q}(\omega)$ tends to the spectral distribution of $q(1+i\omega)$. As elsewhere, we define the moment generating series
\begin{equation}
    F(z) = \tau(R(z)) , \quad \textrm{for} \quad 
    R(z) = (1-z q)^{-1} . 
\end{equation}
Our goal is, as in preceding sections, to obtain a closed functional equation for $F$.

Our approach will be to view $q$ as the solution to the equation
\begin{equation}
    (\zeta-c) q (\zeta^{\ast}-c^{\ast}) = 1,
\end{equation}
and expand this equation in two ways. Defining the shifted circular element $a = \zeta-c$, we can write
\begin{equation}
    a q = (a^{\ast})^{-1} \quad \textrm{and} \quad q a^{\ast} = a^{-1}.
\end{equation}
Tracing both equations against $R$ and using the definition of $a$, this gives
\begin{align}
    \zeta M(z) - \tau( c q R ) &= \tau( (a^{\ast})^{-1} R ),\\ 
    \zeta^{\ast} M(z) - \tau(c^{\ast} q R) &= \tau( a^{-1} R ),
\end{align}
where we let $M(z) = \tau(q R) = \frac{1}{z} (F(z)-1)$ for brevity. We now integrate by parts, using the fact that for $c^{\ast}$ the relevant free difference is $\partial_{c}$. By differentiating the functional equation, 
\begin{equation}
    \partial_{c^{\ast}} q = q \otimes (a^{\ast})^{-1} \quad \mathrm{and} \quad 
    \partial_{c} q = a^{-1} \otimes q.
\end{equation}
Using this result, we then find that
\begin{align}
    \partial_{c^{\ast}} R &= z (R\otimes 1) \partial_{c^{\ast}} q (1 \otimes R) = z R q \otimes (a^{\ast})^{-1} R,
    \\
    \partial_{c} R &= z (R\otimes 1) \partial_{c} q (1 \otimes R) = z R a^{-1} \otimes q R.
\end{align}
Then, by the Leibniz rule, 
\begin{align}
    \partial_{c^{\ast}}(qR) 
    &= (\partial_{c^{\ast}} q) (1 \otimes R) + (q \otimes 1) (\partial_{c^{\ast}} R)
    \\
    &= (q + z q R q) \otimes (a^{\ast})^{-1} R
    \\
    &= q R \otimes (a^{\ast})^{-1} R,
\end{align}
as $q + z q R q = q R$. Similarly, using the fact that $q$ and $R$ commute, 
\begin{equation}
    \partial_{c}(qR)
    = \partial_{c}(Rq) = Ra^{-1} \otimes q R.
\end{equation}
Thus, using the integration by parts identities
\begin{align}
    \tau(c P) &= g (\tau \otimes \tau) \partial_{c^{\ast}} P\\
    \tau(c^{\ast} P) &= g (\tau \otimes \tau) \partial_{c} P,
\end{align}
we have the two equations
\begin{align}
    \zeta M(z) - g M(z) \tau( (a^{\ast})^{-1} R)  &= \tau( (a^{\ast})^{-1} R ) ,
    \\
    \zeta^{\ast} M(z) - g M(z) \tau(a^{-1} R) &= \tau( a^{-1} R ) ,
\end{align}
or
\begin{align}
    \tau( (a^{\ast})^{-1} R) &= \frac{\zeta M(z)}{1 + g M(z)} 
    \\
    \tau( a^{-1} R) &= \frac{\zeta^{\ast} M(z)}{1 + g M(z)}.
\end{align}
To close these equations, we relate $\tau(a^{-1} R)$ and $\tau((a^{\ast})^{-1} R)$ to $F(z)$ in another way. Starting from the defining property that $a a^{-1} = 1$, we expand the definition of $a$ and trace against $R$ to obtain
\begin{equation}
    F(z) = \zeta \tau( a^{-1} R ) - \tau( c a^{-1} R).
\end{equation}
We now integrate by parts, using the fact that $a^{-1}$ is a function only of $c$ and thus has $\partial_{c^{\ast}} a^{-1} = 0$:
\begin{align}
    \tau( c a^{-1} R) 
    &= g (\tau \otimes \tau) (a^{-1}\otimes 1) \partial_{c^{\ast}} R
    \\
    &= g z (\tau \otimes \tau) (a^{-1} R q \otimes (a^{\ast})^{-1} R)
    \\
    &= g z \tau(a^{-1} R q) \tau( (a^{\ast})^{-1} R)
    \\
    &= g [\tau(a^{-1} R) - \tau(a^{-1})] \tau( (a^{\ast})^{-1} R) 
\end{align}
as $R q = \frac{1}{z}(R-1)$. Now, as $\zeta$ is outside the spectral radius of $c$,
\begin{equation}
    \tau(a^{-1}) = \tau((\zeta-c)^{-1}) = \frac{1}{\zeta},
\end{equation}
so we then find that
\begin{equation}
    F(z) = \zeta \tau( a^{-1} R ) - g \left[\tau(a^{-1} R) - \frac{1}{\zeta}\right] \tau( (a^{\ast})^{-1} R) . 
\end{equation}
Substituting the expressions for $\tau(a^{-1} R)$ and $\tau((a^{\ast})^{-1} R)$ that we found before into this equation and algebraically simplifying the result, this gives
\begin{equation}
    F(z) = \frac{(|\zeta|^2 + g) M(z) + g^2 M(z)^2}{(1+g M(z))^2}
\end{equation}
Recalling that $F(z) = 1 + z M(z)$, we can simplify this into a compact cubic equation for $M(z)$:
\begin{equation}
    1 + g M(z) = |\zeta|^2 M(z) - z M(z) (1 + g M(z))^2.
\end{equation}

We now want to use this to solve for the spectral density. To do so, we define $y = - g M(1/\lambda)$, and observe that the Stieltjes transform of the $\zeta$-dependent spectral density is given by 
\begin{equation}
    G(s) = \frac{1}{s} - \frac{y(s)}{g s^2} 
\end{equation}
as $F(z) = 1 + z M(z)$ and $G(s) = F(1/s)/s$. Then, by Stieltjes inversion, the density is
\begin{equation}
    \rho_{\zeta}(\lambda) = -\lim_{\epsilon \downarrow 0} \frac{1}{\pi g \lambda^2} \Im y(\lambda-i\epsilon). 
\end{equation}

The cubic equation for $M(z)$ implies that $y$ is a root of the cubic
\begin{equation}
    y^{3} - 2 y^2 + (1 - (|\zeta|^2-g)\lambda) y - g \lambda = 0.
\end{equation}
This is a cubic with real coefficients, so a root with negative imaginary part exists when the discriminant 
\begin{equation}
    \lambda [4 (|\zeta|^2-g)^3 \lambda^2 - (8 |\zeta|^4 + 20 |\zeta|^2 g - g^2) \lambda + 4 |\zeta|^2]
\end{equation}
is negative. This is possible when $\lambda_{-} < \lambda < \lambda_{+}$, where
\begin{equation}
    \lambda_{\pm} = \frac{2|\zeta|^4 + 5 g |\zeta|^2 - \frac{1}{4} g^2 \pm \frac{1}{4}\sqrt{g} (8|\zeta|^2+g)^{3/2}}{2 (|\zeta|^2-g)^3}
\end{equation}
are the roots of the discriminant for $\lambda>0$. Applying Cardano's formula, we find that in this regime
\begin{equation}
    -\Im y = \frac{3^{1/6}}{2}(\Delta_{+}^{1/3} - \Delta_{-}^{1/3}),
\end{equation}
where
\begin{equation}
\begin{split}
    \Delta_{\pm} &= \left(|\zeta|^2 + \frac{g}{2}\right) \lambda - \frac{1}{9} \\&\quad \pm   \sqrt{\frac{(|\zeta|^2-g)^{3} \lambda (\lambda-\lambda_{-})(\lambda_{+}-\lambda)}{3}} .
\end{split}
\end{equation}
We thus conclude at last that the density of eigenvalues of $\mathbf{Q}(\omega)$ is 
\begin{equation}
    \rho_{\omega}(\lambda) = \frac{3^{1/6}}{2 \pi g \lambda^2}  (\Delta_{+}^{1/3} - \Delta_{-}^{1/3}) \mathbf{1}_{\lambda_{-} < \lambda < \lambda_{+}},
\end{equation}
where we take real cube roots, and we must evaluate $\Delta_{\pm}$ at $|\zeta|^2 = 1 + \omega^2$. 

Accounting for differences in notation, this is precisely the density found by \citet{hu2022spectrum}. \citet{hu2022spectrum} obtained this result using the replica trick; a recent paper from \citet{tian2026diversity} used an alternative block-resolvent free probability approach to extend this result to the case in which $Q(\omega)$ is deformed by inserting a diagonal matrix of gains between the two resolvent factors. 

\section{Numerical methods}\label{sec:numerics}

\begin{figure}[t]
    \centering
    \includegraphics[width=3.3in]{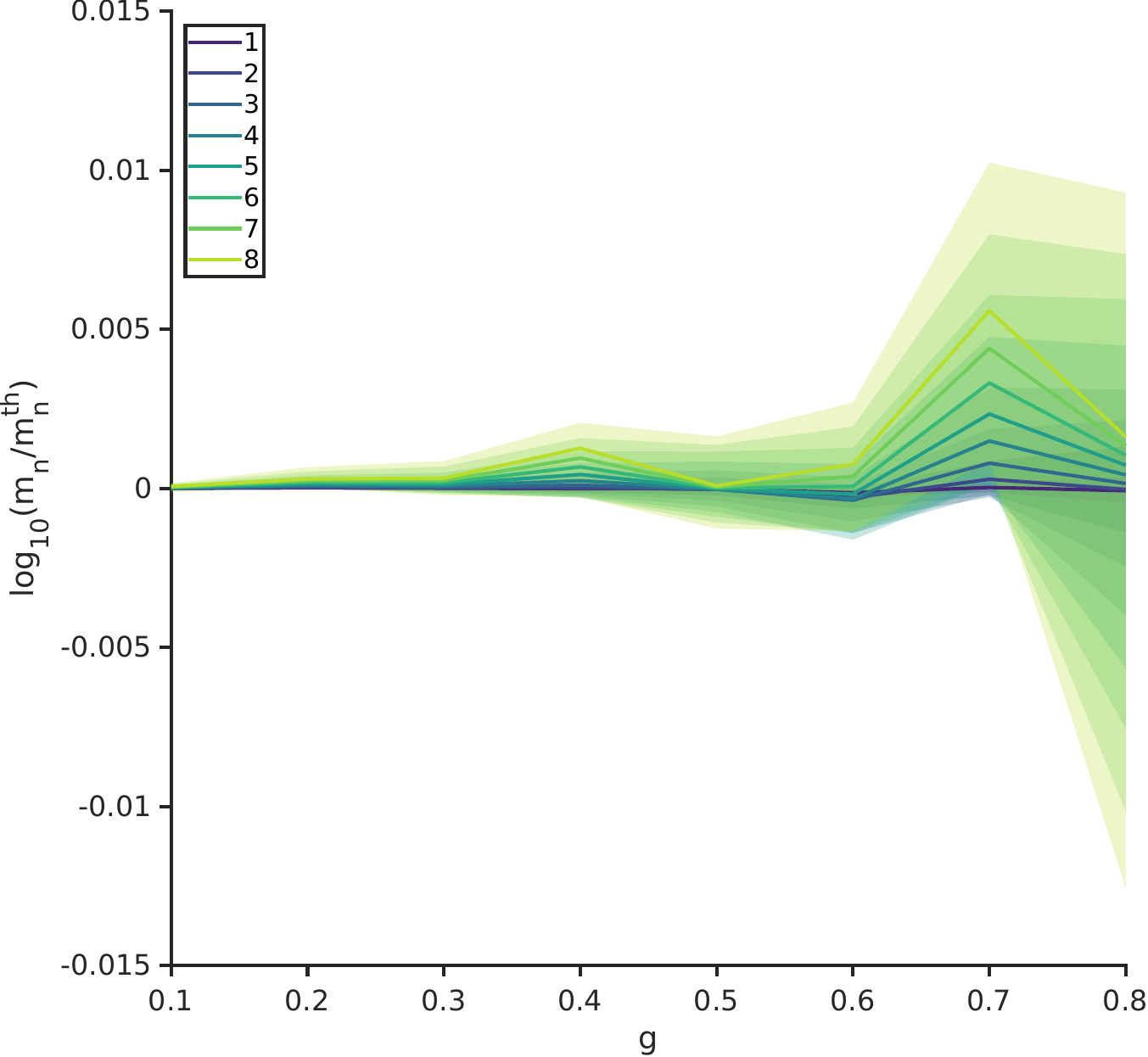}
    \caption{Comparison of empirical and predicted spectral moments across gains $g$. Empirical moments are estimated based on 25 realizations at $N=4096$. Shaded patches show 95\% confidence intervals on the log-relative-error $\log_{10}(\hat{m}_{n}/m_{n}^{th})$ estimated using bootstrapping. }
    \label{fig:moment_comparison}
\end{figure}

Here, we summarize the numerical methods used to generate the figures. The code used to generate Figures \ref{fig:density_comparison} and \ref{fig:tail_collapse} was implemented in Python 3.10, using the NumPy and SciPy libraries \cite{scipy2020}. The code used to generate Figure \ref{fig:moment_comparison} was implemented in Matlab R2024b. All simulations---whether implemented in Python or Matlab---were run on a Dell Precision 7960 desktop workstation equipped with an Intel Xeon(R) w5-3525 processor and 128 GB of RAM. 

To generate the theoretical predictions for the spectral density in Figure~\ref{fig:density_comparison}, we use an approximate numerical scheme to solve the functional equation \eqref{eqn:functional_eqn} and evaluate the inverse Stieltjes transform. To numerically approximate the solution to the functional equation, fix a point $s$ at which we want to evaluate the Stieltjes transform $G(s)=F_{d}(1/s)/s$, and then set $z_0=1/s$. Define the iterates
\begin{equation}
    z_{m+1} = g z_{m} F_{d}(z_{m}),
    \quad \textrm{so} \quad 
    F_{d}(z_{m}) = \frac{1}{g} \frac{z_{m+1}}{z_m} . 
\end{equation}
Then, $F_{d}(z_{m+1}) = F_{d}(g z_{m} F_{d}(z_{m}))$. Using the functional equation $(1-z) F_{d}(z) = F_{d}(g z F_{d}(z))$, we thus find that
\begin{equation}
    \frac{1}{g} \frac{z_{m+2}}{z_{m+1}} = F_{d}(z_{m}) = (1-z_{m}) \frac{1}{g} \frac{z_{m+1}}{z_{m}}.
\end{equation}
Thus, $z_{m}$ obeys the two-term recurrence
\begin{equation}
    z_{m+2} = \frac{(1-z_{m}) z_{m+1}^{2}}{z_{m}}. 
\end{equation}
To fix a boundary condition, we use the fact that $F_{d}(z) \to 1$ as $z \to 0$, which allows us to solve backwards from $z_{M}$ near zero for some large $M$ to $z_{0}$. This limiting behavior implies that
\begin{equation}
    \frac{z_{m+1}}{z_{m}} \to g
\end{equation}
as $m \to \infty$ if the sequence $z_{m} \to 0$. Thus, we set some large cutoff $M$ (in our simulations we find that $M=75$ is large enough to obtain a good match to experiments, though we have not made an effort to optimize this), and set $z_{M} = \alpha g^{M}$ and $z_{M+1} = \alpha g^{M+1}$ for some amplitude $\alpha$. Then, we run the recurrence backwards from $m=M$ to $m=0$ using the expression
\begin{equation}
    z_{m} = \frac{z_{m+1}^2}{z_{m+1}^2+z_{m+2}},
\end{equation}
and then fix $\alpha$ by matching the initial condition $z_{0}=1/s$.

Given this approximate solution, we then recover the Stieltjes transform by using the fact that
\begin{equation}
    G(s) = \frac{1}{s} F_{d}\left(\frac{1}{s}\right) = z_{0} F_{d}(z_{0}) = \frac{1}{g} z_{1}.
\end{equation}
To estimate the spectral density at the point $\lambda$, we set $s = \lambda - i \epsilon$ for some fixed value of $\epsilon$, solve for $z_1$ at each point, and then take $\Im z_1/\pi g$ as our estimate of the density. We choose $\epsilon$ large enough such that the iteration is stable even as $\Re(z_m)$ enters the spectral support; in practice we take $\epsilon = 0.06$. 

We use the same iterative scheme to approximately solve for the upper spectral edge $\lambda_+$. Instead of enforcing $z_0=1/s$, we instead maximize $z_{0}(\alpha)$ with respect to $\alpha$ using \texttt{scipy.optimize.minimize\_scalar}, and then estimate $\lambda_{+} = 1/\max_{\alpha>0} z_{0}(\alpha)$. This gives us a rough estimate of the location where the branch of $F_{d}$ outside the support ceases to become locally invertible. 

To obtain a more quantitative comparison between the empirical and predicted spectral distributions, in Figure~\ref{fig:moment_comparison} we consider the first eight moments across varying values of $g$. To produce the theoretical predictions, we solve the moment recurrence \eqref{eqn:moment_recurrence} by hand for the first few orders, and then using SymPy up to $n=8$. The expressions for the moments are rational functions of $g$ whose precise forms are not illuminating; we therefore do not list them here, and direct the reader to the GitHub for the SymPy code, a draft version of which was generated by GPT 5.5 Pro. These predictions are then passed into a Matlab script which compares them against the result of numerically estimating the spectral moments across a range of values of $g$; this forms the basis for Figure~\ref{fig:moment_comparison}. We restrict our attention to $g \leq 0.8$ as the theory predicts that $m_{8}(g=0.8)$ is on the order of $10^{11}$; larger values and higher moments are even larger, and numerical stability becomes an issue. Moreover, high-order moments at large $g$ are dominated by rare large eigenvalues.

This is therefore a fully independent numerical check, as it relies on a different method for computing the empirical spectrum than the Python code used elsewhere. Here, as in Figure~\ref{fig:tail_collapse}, we enforce a stability threshold on each realization of $\mathbf{J}$ so that the computation of $\mathbf{\Sigma}_{d}$ is numerically stable. Concretely, in code we generate new realizations of $\mathbf{J}$ until the condition $\rho(\mathbf{J}) < 1-\varepsilon$ is satisfied, for a stability threshold $\varepsilon = 10^{-3}$. In practice, we found that one attempt was sufficient for all of the realizations at $N=4096$ used in Figure~\ref{fig:moment_comparison}. There, we show the log-ratio $\log_{10}(\hat{m}_{n}/m_{n}^{th})$ between the measured and predicted moments across $g$; from this we see that the ratiometric error is at most around $10^{0.015} \simeq 1.035$ for $g=0.8$, where the raw value of the moment is around $10^{11}$.

\bibliography{refs}

\end{document}